\documentclass[a4paper,fleqn]{cas-dc}

\usepackage{cite}

\usepackage[numbers]{natbib}
\usepackage{siunitx}
\usepackage{lineno}

\usepackage{placeins}

\usepackage{tabularx}
\usepackage{subcaption}

\def\tsc#1{\csdef{#1}{\textsc{\lowercase{#1}}\xspace}}
\tsc{WGM}
\tsc{QE}

\begin{document}

\let\WriteBookmarks\relax
\def\floatpagepagefraction{1}
\def\textpagefraction{.001}

\shorttitle{}    

\shortauthors{}  

\title [mode = title]{High-Speed Semi-FE Readout Module for ATLAS MDT at HL-LHC: Design and Production-Level Characterization}  


\tnotetext[1]{Corresponding author: huxueye@ustc.edu.cn} 

%

\author[1]{Yao Teng}
\author[2]{Xueye Hu$^*$}
\author[1]{Yuxiang Guo}
\author[1]{Shaghayegh Emami}
\author[1]{Thomas Baer}
\author[1]{Thomas Schwarz}
\author[1]{Junjie Zhu}
\author[1]{Bing Zhou}






\affiliation[1]{organization={Department of Physics, University of Michigan},
            addressline={450 Church Street}, 
            city={Ann Arbor},
            postcode={48109},
            state={Michigan},
            country={USA}}

\affiliation[2]{organization={Department of Modern Physics, University of Science and Technology of China},
            addressline={96 Jinzhai Road},
            city={Hefei},
            postcode={230026},
            state={Anhui},
            country={China}}

\begin{abstract}
The High-Luminosity upgrade of the Large Hadron Collider (HL-LHC) introduces increased demands on the ATLAS Muon Spectrometer, particularly in terms of data throughput, timing distribution and system reliability. The Phase-II Chamber Service Module (CSM) is a key component of the upgraded Monitored Drift Tube (MDT) trigger and readout system, providing a high-speed interface between the front-end electronics and the backend systems. 
This paper describes the design and implementation of the Phase-II CSM, together with its validation. The results show that the CSM supports two independent optical uplinks,
each operating at a line rate of
\SI{10.24}{Gbps}, together with clock distribution and
slow control in the expected operating environment. Integration with small-diameter MDT (sMDT) chambers and tests with the prototype L0MDT trigger system are also presented. The CSM boards are now in production and will be used for installation and integration during the upcoming LHC Long Shutdown. 
\end{abstract}


\begin{highlights}
\item 
\item 
\item 
\end{highlights}


\maketitle


\section{Introduction}\label{}

The ATLAS detector is a general-purpose particle detector at the Large Hadron Collider (LHC) and is designed to detect the products of proton-proton collisions at multi-TeV energy scales. The data collected enables physicists to perform high-precision tests of the Standard Model (SM) and search for new physics beyond the SM \cite{atlas2013physics}.

The main goal of the High-Luminosity LHC (HL-LHC) upgrade is to increase the LHC's peak luminosity to 5-\qty{7e34}{cm^{-2}s^{-1}} and to deliver an integrated luminosity of up to 3-\SI{4}{ab^{-1}} over its lifetime \cite{aberle2020high}. However, this upgrade will result in an average 200 simultaneous collisions per bunch crossing (25 ns), thereby dramatically increasing trigger rates, data throughput, and radiation exposure. Substantial upgrades to the ATLAS detector, electronics, trigger and data acquisition (TDAQ) systems are needed. In particular, ATLAS will implement a hardware-based first stage in its trigger system, known as Level-0 (L0), which will have an output trigger rate of 1 MHz and a latency of \SI{10}{\us} \cite{atlas2022technical}.

The ATLAS Muon Spectrometer (MS) employs a variety of detector technologies for muon triggering, identification, and momentum measurement. Precise tracking is primarily provided by the Monitored Drift Tube (MDT) chambers, while fast triggering is achieved with Resistive Plate Chambers (RPCs) in the barrel region and Thin Gap Chambers (TGCs) in the endcap \cite{izumiyama2025overview}. 

Maintaining sensitivity to low transverse momentum ($p_{\mathrm{T}}$) muons without increasing the trigger threshold is essential for HL-LHC runs. However, the relatively coarse momentum resolution of the RPC and TGC detectors can cause some low-$p_{\mathrm{T}}$ muons to be reconstructed as higher-$p_{\mathrm{T}}$ candidates~\cite{policicchio2019phase}. The first-level muon trigger rate is consequently dominated by low-$p_{\mathrm{T}}$ muons below the nominal trigger threshold that are accepted because of the limited momentum resolution of the trigger chambers~\cite{Nowak2016MDTTrigger}. With the HL-LHC upgrade, the expected \SI{10}{\us} latency and \SI{1}{MHz} trigger rate at L0 open the possibility of using the MDT detector for triggering \cite{atlas2022technical}. The MDT's precise measurement of muon hit positions will improve momentum resolution at the trigger level and reduce the trigger rate. 

The ATLAS MDT detector consists primarily of chambers using \SI{3}{cm} diameter drift tubes, while about 150 out of approximately 1150 MDT chambers use \SI{1.5}{cm} diameter tubes. When a muon passes through a tube, it ionizes the Ar: CO$_2$ (93:7) gas mixture, and the resulting electrons drift toward the central wire \cite{TheATLASCollaboration_2008}. 
The signal from the central wire is processed by an
Amplifier/Shaper/Discriminator (ASD) ASIC~\cite{Kroha:2016fid, DeMatteis:2017xky},
which determines the signal arrival time from the discriminator threshold
crossing. The ASD also incorporates a Wilkinson charge-to-time converter,
which encodes the collected signal charge into the width of the discriminator
output pulse~\cite{Abovyan2019ASD2}. The TDC ASIC records both the leading
and trailing edges of this pulse, providing the hit arrival time and the pulse
width, respectively~\cite{Wang:2017jnd, Liang:2019weg,
guo2023development}. The pulse width provides a measure of the signal charge
and can be used for time-slewing correction and detector monitoring.
Each ASD ASIC handles eight tubes, while each TDC receives the discriminated
signals from three ASDs. A mezzanine card containing three ASDs and one TDC
therefore serves 24 tubes~\cite{arai2008atlas}.

The current MDT readout system operates at a trigger rate of $\sim$100 kHz and \SI{2.5}{\us} latency. Hits are stored in buffered memories of the TDC waiting for the ATLAS first-level trigger acceptance signal. The maximum drift time of a \SI{30}{\mm} MDT tube is approximately \SIrange{700}{750}{\ns}. In the present triggered readout, an accepted trigger causes the TDC to select hits within a trigger-matching window of approximately \SI{1.3}{\us}. At an L0 trigger rate approaching \SI{1}{\MHz}, corresponding to an average interval of about \SI{1}{\us} between consecutive trigger accepts, neighboring matching windows can overlap, allowing the same buffered hit to be selected by more than one trigger. Triggerless streaming eliminates this trigger-matching duplication at the front-end by transmitting each digitized MDT hit once to the off-detector electronics, where trigger association is subsequently performed~\cite{guo2023development}.  

\begin{figure*}[t]
    \centering
    \includegraphics[width=0.95\linewidth]{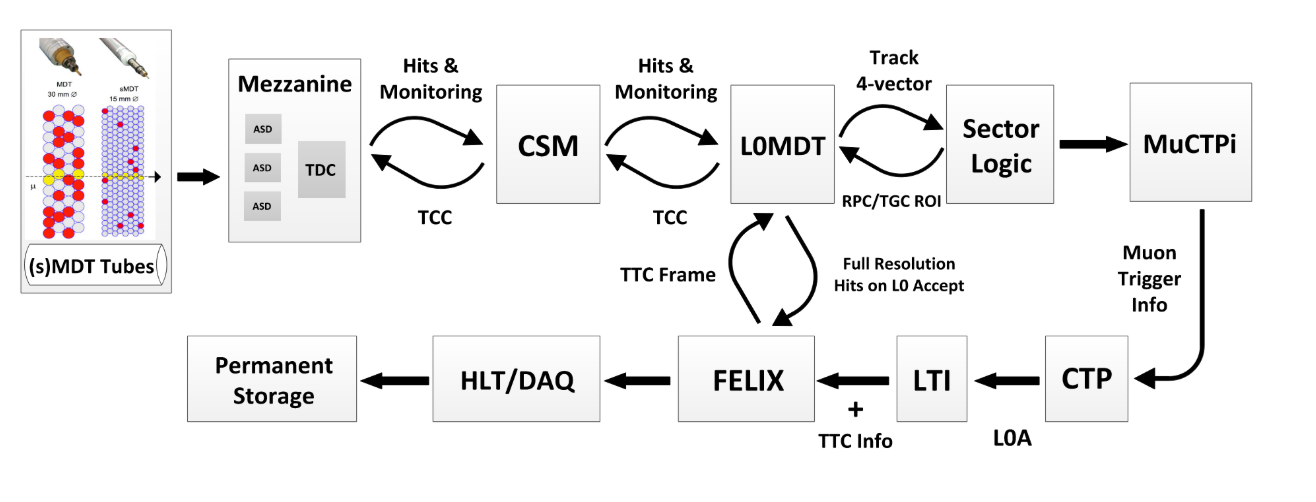}
    \caption{Data flow architecture of the Phase-II ATLAS Muon MDT trigger and readout system. The CSM sits between the L0MDT processor and the front-end mezzanine
cards (ASD and TDC), aggregating MDT hit data and distributing timing,
control, and configuration (TCC) signals. Trigger candidate information
is forwarded through the Sector Logic board to the Muon-to-Central
Trigger Processor Interface (MuCTPi) for selection and overlap removal,
and then sent to the Central Trigger Processor (CTP). The CTP issues a
Level-0 Accept (L0A) signal based on the global trigger logic. The L0A,
together with Timing, Trigger, and Control (TTC) information, is
distributed by the Local Trigger Interface (LTI) through FELIX.
Upon receiving the L0A, FELIX requests full-resolution hit data from
the L0MDT for accepted events, which are routed to the High-Level
Trigger (HLT) and data acquisition (DAQ) pipeline for further processing
and permanent storage.}
    \label{fig:csm_architecture}
\end{figure*}

Figure~\ref{fig:csm_architecture} shows the architecture of the proposed MDT TDAQ system for HL-LHC runs\cite{Hu:2886750, CERN-LHCC-2017-017}. All hits detected by the ASD and TDC ASICs are transmitted to the Chamber Service Module (CSM), which aggregates the complete hit-data stream from up to 20
front-end mezzanine cards and transmits it to the backend through
high-speed optical links. Data are then sent via an optical module to the L0MDT board (also known as the MDT Trigger Processor), where relevant hits are extracted. 

To enable triggerless operation of the HL-LHC, we designed a new generation of CSM boards featuring high bandwidth, fixed latency, and enhanced radiation tolerance. In this work, we detail their design, testing procedures, and performance validation.

\section{CSM design}

In HL-LHC runs, all existing front-end and backend electronics will be
replaced with newly developed systems. Approximately 18,000 mezzanine
cards with redesigned ASD and TDC chips, together with 1,104 CSM boards,
will be installed~\cite{policicchio2019phase}. Each front-end mezzanine
card provides two TDC serial links operating at
\SI{320}{Mbps}~\cite{CERN-LHCC-2017-017}. For a CSM
serving up to 20 mezzanine cards, the 40 TDC links therefore correspond
to an aggregate MDT hit-data input line rate of
\SI{12.8}{Gbps}. Since the upgraded system operates in
triggerless mode, the CSM must continuously accommodate this hit-data
throughput. This requirement is a primary driver of the CSM readout
architecture.

\subsection{Hardware Design}

\begin{figure*}[t]
    \centering
    \includegraphics[width=0.85\linewidth]{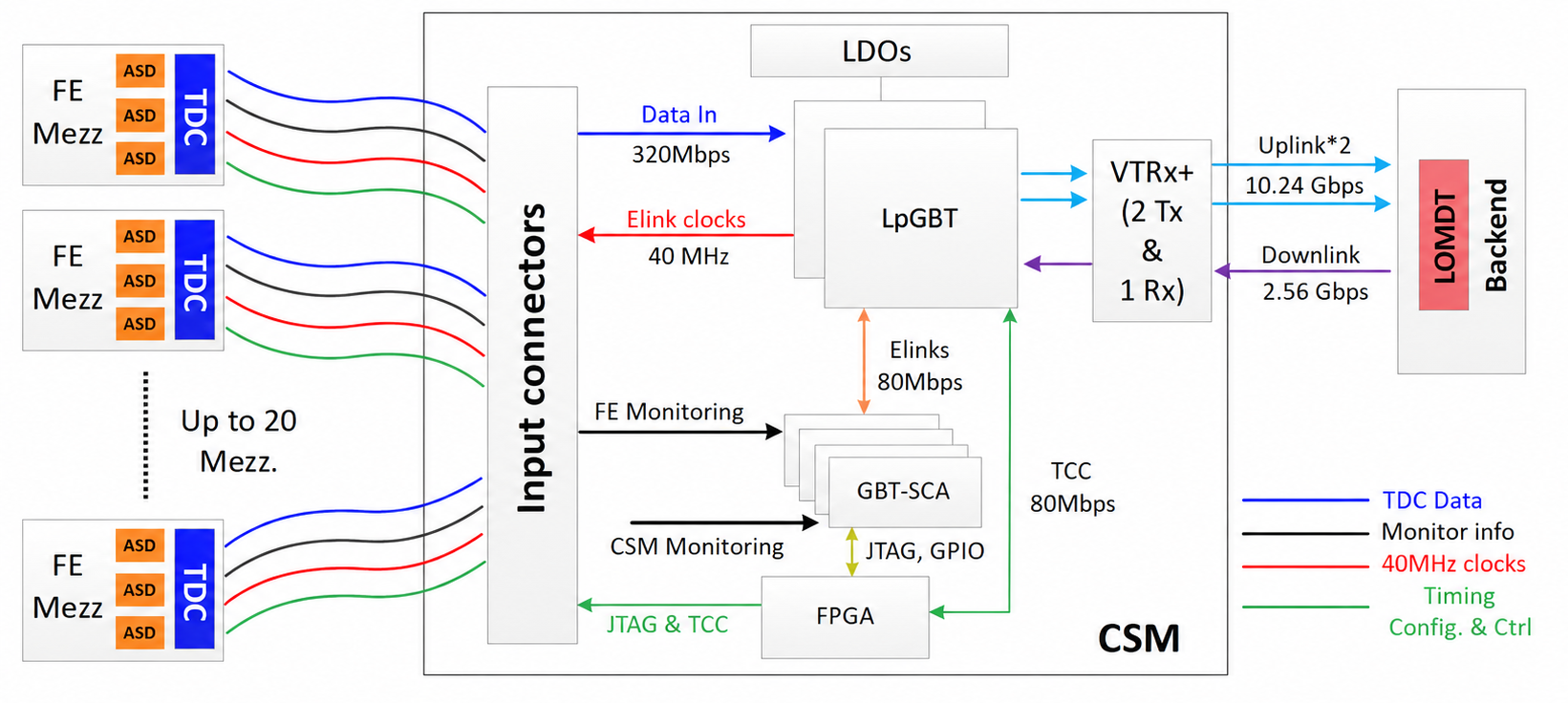}
    \caption{Functional architecture of the CSM. The board interfaces with
    up to 20 front-end mezzanine cards and communicates with the backend
    through two \SI{10.24}{Gbps} uplinks and one
    \SI{2.56}{Gbps} downlink via the VTRx+ optical module.
    The low-dropout regulators (LDOs) provide the required on-board power
    rails. Control and monitoring are handled by the GBT-SCA ASICs and FPGA
    logic. The GBT-SCA data are transmitted at
    \SI{80}{Mbps}, while the corresponding lpGBT E-link
    interface operates at \SI{320}{Mbps}. Each SCA data bit
    therefore spans four lpGBT E-link bit periods.}
    \label{fig:csm_hardware_architecture}
\end{figure*}

The primary role of the CSM is to aggregate the full hit-data stream
from up to 20 front-end mezzanine cards and transmit it to the backend
through a smaller number of high-speed optical links. The CSM does not reduce the hit-data bandwidth by filtering. Instead,
it consolidates the large number of front-end serial links into two
optical uplinks.
In parallel, the CSM provides timing, control, and configuration
distribution, as well as configuration and monitoring of the front-end
electronics. Figure~\ref{fig:csm_hardware_architecture} shows the
overall hardware architecture and signal connections
~\cite{Hu:2886750, moreira2024lpgbt}.

In the Phase-II design, the mezzanine configuration function, previously handled through the Embedded Local
Monitor Board (ELMB)~\cite{arai2008atlas}, is now integrated into the CSM functionality. The CSM can therefore configure the mezzanine cards directly, which optimizes the system architecture. As shown in Figure \ref{fig:csm_hardware_architecture}, the new CSM incorporates four GBT-Slow Control Adapter (GBT-SCA) ASICs~\cite{gabrielli2009gbt} for slow control and mezzanine card monitoring. Each SCA provides 31 ADC channels, of which 30 are externally accessible
and used to monitor the analog supply voltages of the front-end circuitry,
the digital supply voltages, and temperatures on the mezzanine cards. The remaining channel is reserved for internal monitoring within the GBT-SCA. The GBT-SCA also includes a Joint Test Action Group (JTAG) interface, which is used to configure the Field-Programmable Gate Array (FPGA) remotely and provides a reliable path for firmware updates and recovery~\cite{gabrielli2009gbt}. 

The JTAG path used to configure the mezzanine TDCs is separate from
this FPGA-configuration interface. An AMD Artix-7 FPGA implements the
JTAG fanout and dynamically constructs the configuration chain for the
mezzanine cards. The signals are routed through high-density connectors
to the motherboard and then to the individual mezzanine cards through
dedicated JTAG cables. Faulty or disconnected mezzanine cards can be
bypassed so that the remaining cards remain accessible.

The CSM board transmits all hit events without performing on-board
data-package filtering. Hit data, together with control and monitoring signals, are routed directly from mezzanine cards or the GBT-SCA to the Low Power Gigabit Transceiver (lpGBT). Two lpGBT ASICs are
located on each CSM, and each provides an independent optical
uplink operating at a line rate of
\SI{10.24}{Gbps}. Only one lpGBT is equipped with an
optical downlink, operating at
\SI{2.56}{Gbps}. This device is referred to as the
master lpGBT, while the other device, which provides only an uplink, is
referred to as the slave lpGBT. The corresponding link configuration
and bandwidths are summarized in Table~\ref{tab:csm_link_config}.

\begin{table}[t]
    \centering
    \caption{Summary of CSM data-link configuration.}
    \label{tab:csm_link_config}
    \begin{tabular}{lcc}
        \hline
        \textbf{Source / link} & \textbf{Number of links} & \textbf{Rate per link} \\
        \hline
        TDC E-links              & 40 & \SI{320}{Mbps} \\
        GBT-SCA PRI/AUX E-links  & 8  & \SI{320}{Mbps} \\
        Total active E-links     & 48 & \SI{320}{Mbps} \\
        Master lpGBT optical uplink & 1 & \SI{10.24}{Gbps} \\
        Slave lpGBT optical uplink & 1 & \SI{10.24}{Gbps} \\
        Master lpGBT downlink & 1 & \SI{2.56}{Gbps} \\
        \hline
    \end{tabular}
\end{table}

The pre-production CSMs were assembled with lpGBTv1 ASICs for
functional verification. The known issues identified in lpGBTv1 did
not affect the CSM functions validated during pre-production. All
production CSMs use lpGBTv2 ASICs, in which the corresponding design
issues were corrected~\cite{HernandezMontesinos2026lpGBT}.

The uplink traffic consists of the 40 TDC E-links carrying MDT hit data
and eight additional E-links from the four GBT-SCA ASICs.  Since every E-link
operates at \SI{320}{Mbps}, the TDC contribution
corresponds to an input rate of
\SI{12.8}{Gbps}, while the eight GBT-SCA links add
\SI{2.56}{Gbps}. 

Each lpGBT uplink has a physical line rate of
\SI{10.24}{Gbps}. In the FEC5 configuration, each
256-bit uplink frame contains 224 bits (28 8-bit E-link channels) available for user data, giving
a user-data capacity of
\SI{8.96}{Gbps} per lpGBT.
The 48 E-link inputs are distributed between the two lpGBT ASICs (56 available E-link channels) and
transmitted to the backend through two independent optical uplinks.

The slave lpGBT receives the remaining 20 TDC E-links, leaving eight
user-data channels unused. The unused slave lpGBT positions are not dynamically
reassigned. The detailed frame mappings are given in
Tables~\ref{tab:master_uplink} and~\ref{tab:slave_uplink}.

The master lpGBT downlink operates at
\SI{2.56}{\giga\bit\per\second} and carries slow-control data from the
backend to the GBT-SCA ASICs through their E-link interfaces. For remote
FPGA firmware configuration or reconfiguration, the configuration data
are subsequently transferred from the GBT-SCA to the FPGA through the
GBT-SCA JTAG interface. This path is distinct from the JTAG fanout
implemented by the FPGA for configuring the mezzanine TDCs. The downlink
frame mapping is specified in Table~\ref{tab:downlink_format} in
Appendix~\ref{appendix}.

The FPGA uses volatile configuration memory and therefore does not
retain its configuration after a power cycle. The CSM contains no
on-board nonvolatile FPGA configuration memory. After each power
cycle, the FPGA is configured remotely from the backend through the
master lpGBT downlink and the GBT-SCA JTAG interface. Configuration
of one CSM through this path requires approximately
\SI{10}{\second}. The total configuration time for
the installed population therefore depends on the degree of
parallelism implemented in the backend control software.

The CSM recovers the global \SI{40}{\mega\hertz} clock through the
lpGBT and distributes it to all mezzanine cards, providing a common
timing reference for the front-end electronics. The channel-to-channel
clock skew at the mezzanine card connectors was not measured in this
study. Accordingly, the quantitative value reported here is limited
to the lpGBT specification. The device provides fixed and
deterministic link latency and a specified clock output jitter below
\SI{5}{\pico\second} RMS~\cite{moreira2024lpgbt}.

\begin{figure}[t]
    \centering
    \includegraphics[width=\linewidth]{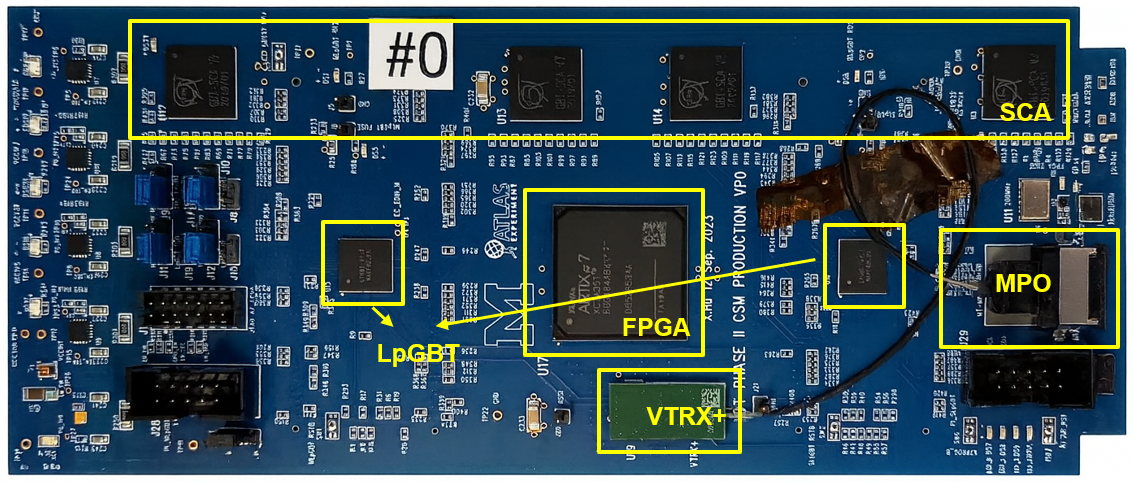}
    \caption{Photograph of a production CSM board. Visible components include the AMD Artix-7 FPGA, lpGBT transceivers, GBT-SCAs, and the VTRx+ optical interface. The label ``\#0'' marking is an identifier and has no functional meaning.}
    \label{fig:csm_board_photo}
\end{figure}

A photograph of the assembled CSM is shown in Figure~\ref{fig:csm_board_photo}. The Printed Circuit Board (PCB) has a size of 81.3~mm~$\times$~194~mm and consists of 16 layers.

\subsection{Firmware Design}

\begin{figure*}[t]
    \centering
    \includegraphics[width=0.88\linewidth]{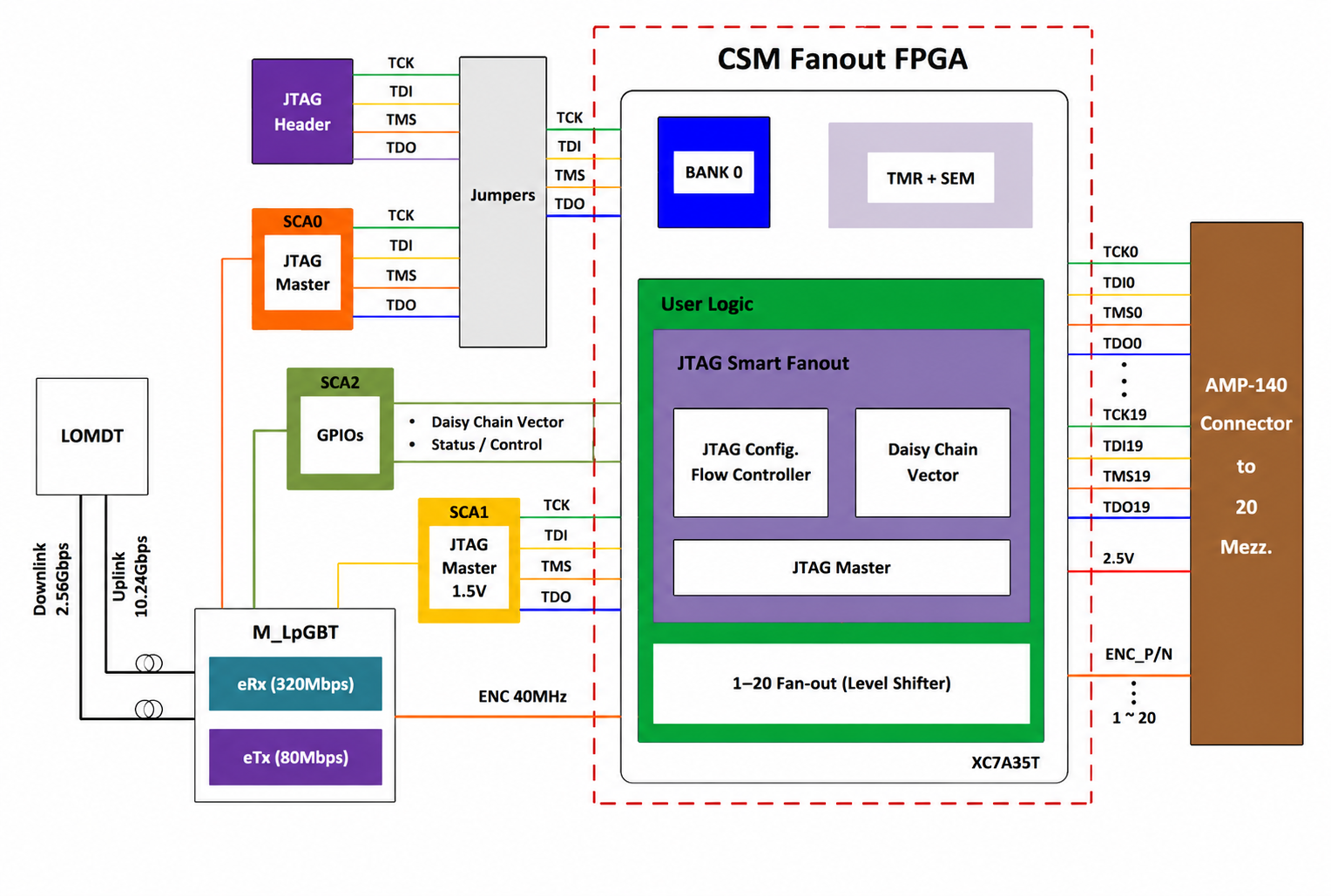}
    \caption{Firmware architecture of the CSM fanout FPGA as well as the interactions between the FPGA, lpGBT and GBT-SCAs. The user logic includes a smart JTAG fanout controller, driven by daisy chain vectors received from the backend system via GBT-SCA GPIOs. A JTAG master coordinates selective configuration of up to 20 mezzanines through a 140-pin high-density connector. TMR and SEM modules provide protection against radiation-induced upsets, while ENC and monitoring signals are synchronized through the lpGBT and GBT-SCA interfaces. GBT-SCA3 is not shown in the diagram as it is not connected to the FPGA and is solely used for mezzanine monitoring via its ADC channels.}
    \label{fig:csm_firmware_architecture}
\end{figure*}

Figure~\ref{fig:csm_firmware_architecture} shows the CSM firmware architecture. In addition it also highlights the interactions between the FPGA, lpGBT and GBT-SCAs, together with the logic used for configuration fanout, clock distribution, and radiation fault protection.

The main function of the firmware is to route JTAG and Enable Control (ENC) signals to the mezzanine cards. During initialization, the firmware checks the connectivity of each mezzanine slot and identifies which cards are present.

Remote FPGA configuration data are received through the optical downlink and transferred to the FPGA through the master GBT-SCA. Configuration data for the front-end mezzanine cards are then distributed by the FPGA through the corresponding GBT-SCA E-links.

To improve tolerance against radiation induced soft errors, the firmware uses a fault tolerant design that includes both logic level and configuration level protection. A Triple Modular Redundancy (TMR) scheme is applied to most of the critical user logic~\cite{lyons1962use}. The logic is replicated three times, and the outputs are combined using majority voting. This approach reduces the impact of Single Event Upsets (SEUs) on FPGA operation.

At the same time, a Soft Error Mitigation (SEM) controller monitors the configuration memory of the FPGA. The SEM uses built-in Error Correction Code (ECC) and Cyclic Redundancy Check (CRC) to detect and correct single-bit errors and double-bit adjacent errors in the Configuration RAM (CRAM)~\cite{amd_sem_controller_pg036}.

Additionally, if an error cannot be corrected by TMR or SEM, the FPGA can be remotely reconfigured from the backend electronics using the same lpGBT/GBT-SCA configuration path. If reconfiguration does not restore correct operation, the CSM can be power cycled and the FPGA subsequently reconfigured.

These mechanisms work together to maintain stable firmware operation in the HL-LHC radiation environment. TMR protects the user logic, and SEM protects the configuration memory. Reconfiguration provides recovery at the FPGA level, while power cycling provides recovery at the system level.

\section{Performance and Reliability Tests}
\subsection{Production Level Tests}\label{sec:standalone}

\begin{figure}[t]
    \centering
    \includegraphics[width=0.95\linewidth]{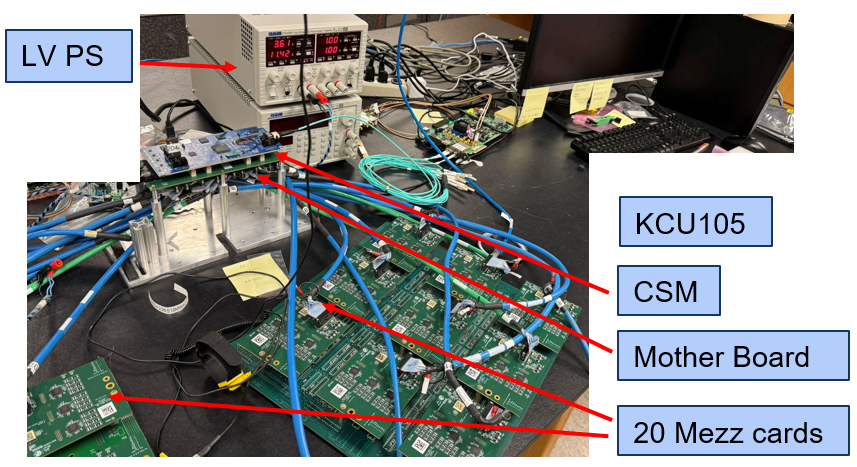}
    \caption{Production level test platform for the CSM. The bench includes power supplies, FPGA controller, custom interface motherboard, and full mezzanine configuration.}
    \label{fig:board_test_setup}
\end{figure}

A dedicated production-level test platform was constructed before integrating the CSM into the full readout system, as shown in Figure~\ref{fig:board_test_setup}. 
The HL-LHC upgrade requires up to 1104 CSM boards, with an additional production margin of 10\%.

The platform consists of a low-voltage power supply, an AMD KCU105 evaluation board as the DAQ system, one CSM, up to 20 mezzanine cards, and a motherboard accommodating the connection between them.

\begin{figure*}[t]
    \centering
    \includegraphics[width=0.95\linewidth]{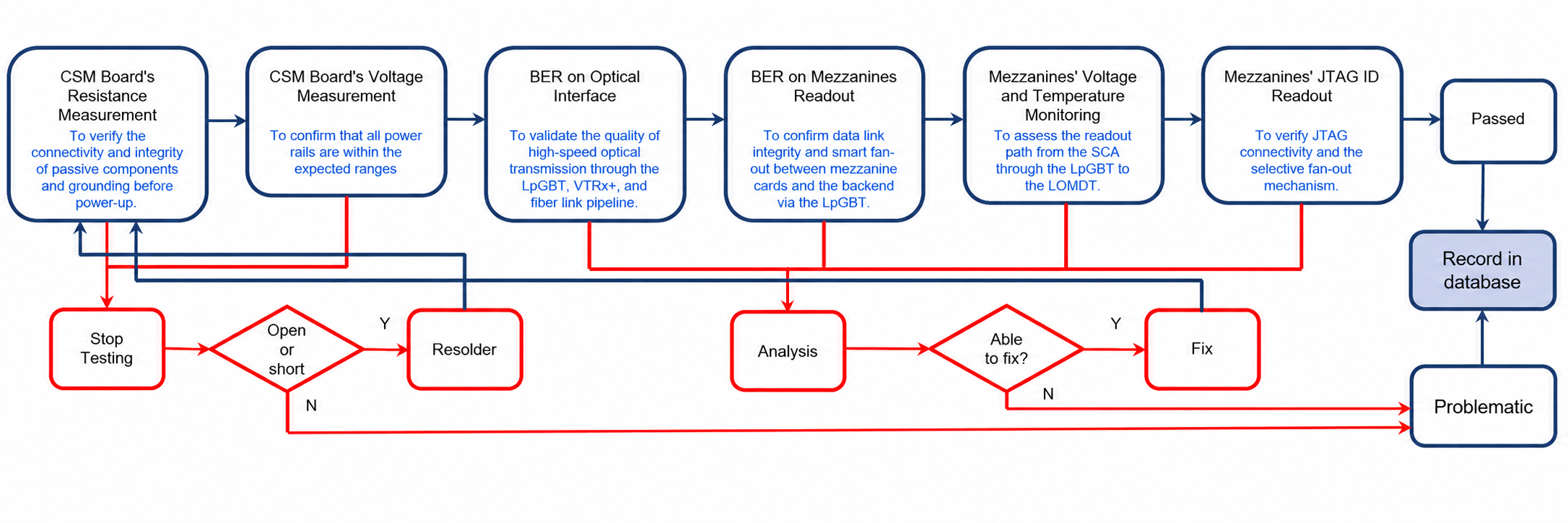}
    \caption{Flowchart summarizing the production level platform test sequence for CSM board validation. Each test step includes a specific purpose. Only boards passing all checks are qualified for system integration.}
    \label{fig:csm_test_flow}
\end{figure*}

The overall test flow is shown in Figure~\ref{fig:csm_test_flow}. 
Each step corresponds to a specific validation purpose. 
All results are recorded in a dedicated database.

Before mass production, a pre-production batch of 50 boards was tested using the same procedure. All 50 pre-production CSMs underwent the complete validation sequence described below.

Resistance measurements were first performed at critical test points to detect shorts or open circuits from possible manufacturing defects. 
These checks help prevent short circuit damage during power-up. 

Supply voltages from the onboard Low Dropout (LDO) regulators were then measured. 
The regulated output voltages were measured relative to their nominal design values. The predefined production acceptance criterion required each output voltage to remain within $\pm5\%$ of its nominal value. Across all 50 boards, the measured deviations were within $\pm2\%$. 
This indicates stable operation of the power distribution system.

The optical link performance was evaluated using a Pseudo-Random Bit
Sequence (PRBS)-31 pattern at the nominal lpGBT uplink line rate of
\SI{10.24}{\giga\bit\per\second}. The complete optical path consisted
of the lpGBT chips, the VTRx+ optical module, a
\SI{1}{\meter} optical fiber, and an SFP+ receiver connected to an AMD
KCU105 evaluation board. A separate scan over alternative transmitter or
receiver equalization settings was not part of this study. Both
optical uplinks of all 50 pre-production CSMs were tested. The test
duration varied among links, with most links tested for approximately
\SI{1}{\hour}. Some runs were shorter because of operational
constraints, while others were extended beyond \SI{1}{\hour}. No bit
errors were observed in any of the PRBS-31 tests.

\begin{figure}[t]
    \centering
    \includegraphics[width=0.85\linewidth]{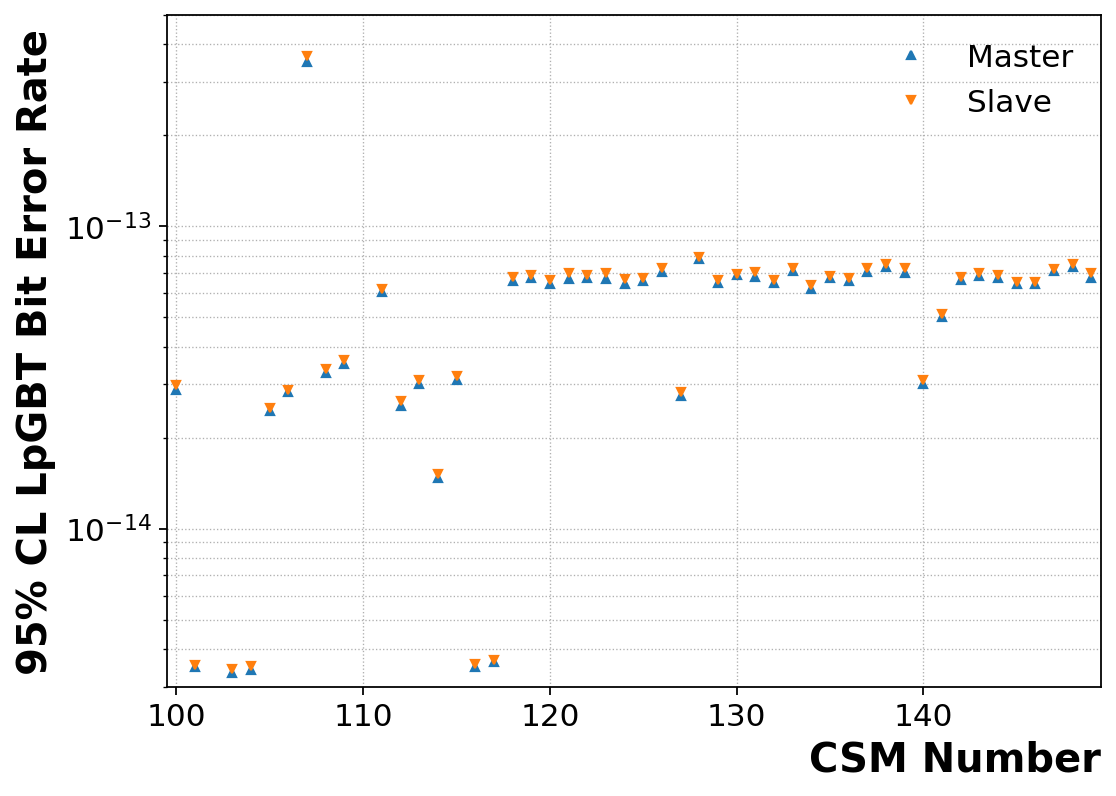}
    \caption{One-sided 95\% confidence upper limits on the BER for the
master and slave lpGBT optical uplinks across the tested CSM units.
No bit errors were observed. The different upper limits result from
variations in test duration, and hence in the number of transmitted
bits \(N_i\), among the individual links.}
    \label{fig:lpGBT_BER}
\end{figure}

Since no errors were observed, the values shown in
Figure~\ref{fig:lpGBT_BER} are not direct BER measurements, but
one-sided upper confidence limits. For a given link \(i\), the upper
limit at confidence level \(C\) is

\begin{equation}
    \mathrm{BER}_{C,i}
    =
    \frac{-\ln(1-C)}{N_i},
\end{equation}

where \(N_i\) is the total number of bits transmitted during the test
of that link. At the fixed line rate used here,
\(N_i = R t_i\), where \(t_i\) is the corresponding test duration.
For example, a \SI{1}{\hour} test at
\SI{10.24}{\giga\bit\per\second} corresponds to
\(N_i \simeq \num{3.69e13}\), giving a 95\% confidence upper limit of

\begin{equation}
    \mathrm{BER}_{95,i}
    \simeq
    \num{8.1e-14}.
\end{equation}

\begin{figure}[t]
    \centering
    \includegraphics[width=0.85\linewidth]{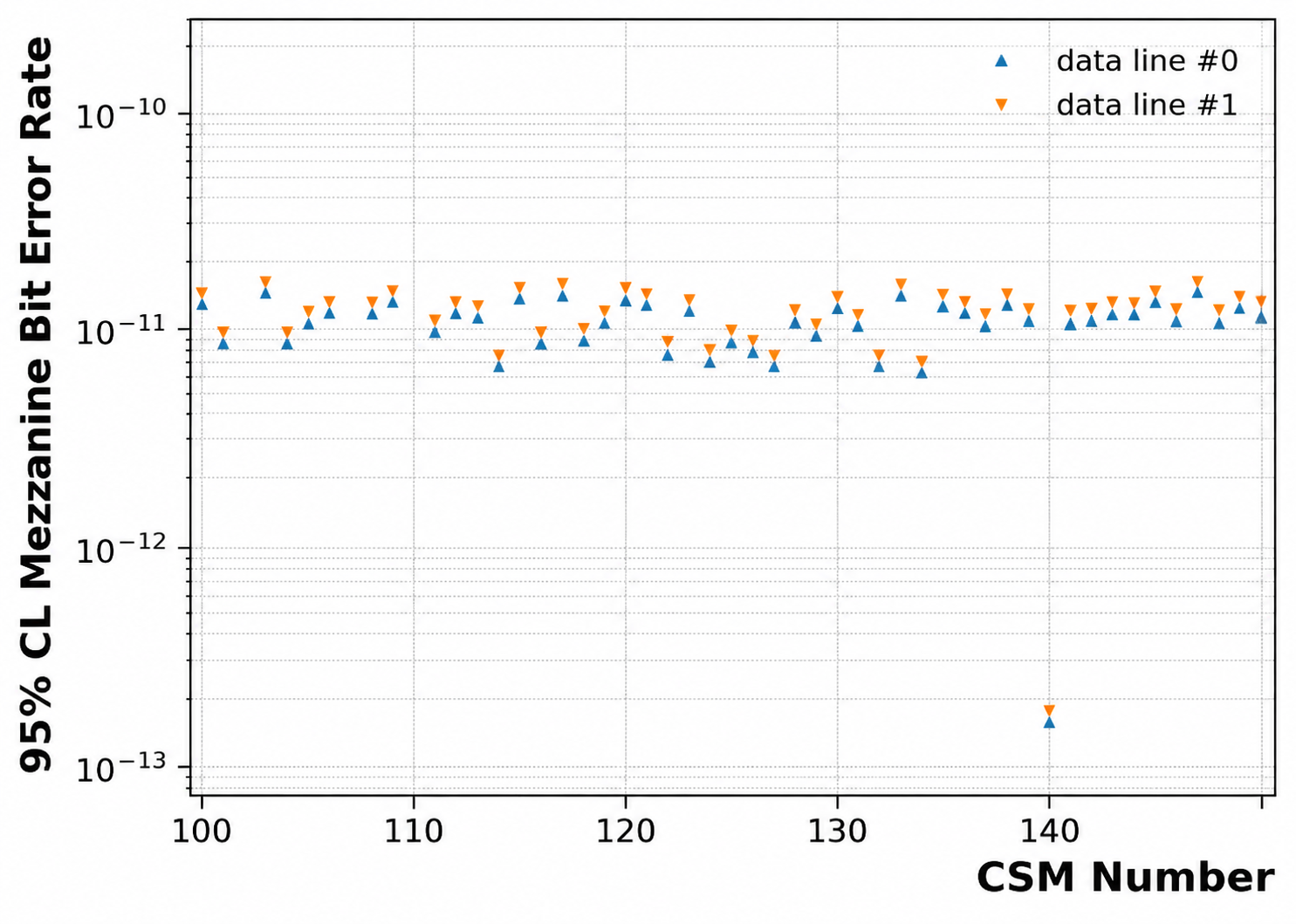}
    \caption{One-sided 95\% confidence upper limits on the BER for the
40 mezzanine-TDC E-link channels of each tested CSM. All channels on
a given CSM were tested simultaneously with common start and stop
times, and no errors were observed.}
    \label{fig:mezz_BER}
\end{figure}

The full readout chain was tested with all 20 mezzanine cards
connected. For each CSM, all 40 TDC E-link channels were exercised
simultaneously with common start and stop times. No bit errors were
observed. Because every channel on a given CSM accumulated the same
number of transmitted bits, the channels have the same one-sided 95\%
confidence upper limit, and one representative value is therefore
shown for each CSM in Figure~\ref{fig:mezz_BER}. All resulting upper
limits are below \num{2e-11}. 

\begin{figure*}[t]
    \centering
    \begin{subfigure}[t]{0.49\textwidth}
        \centering
        \includegraphics[width=\linewidth]{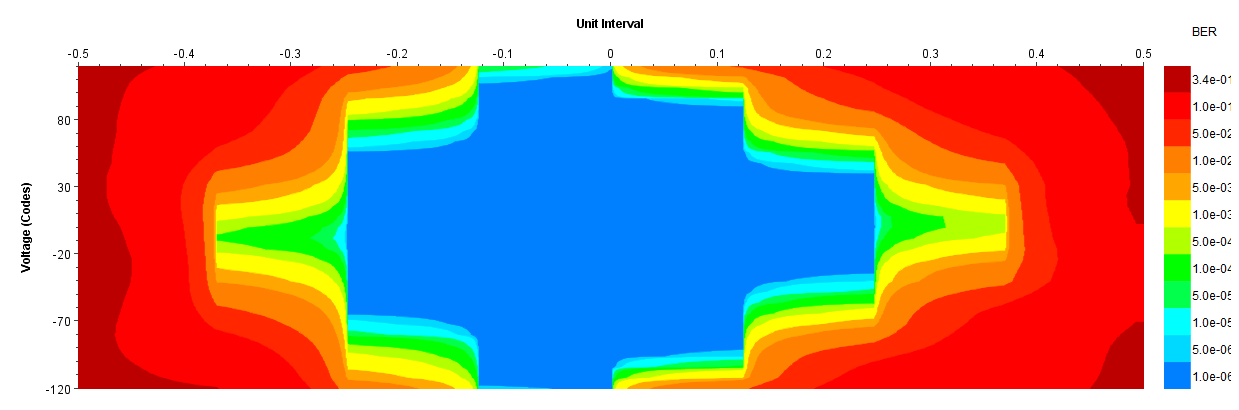}
        \caption{Master lpGBT uplink: open area 6016 (\SI{56}{\percent}).}
        \label{fig:eyescan_master}
    \end{subfigure}
    \hfill
    \begin{subfigure}[t]{0.49\textwidth}
        \centering
        \includegraphics[width=\linewidth]{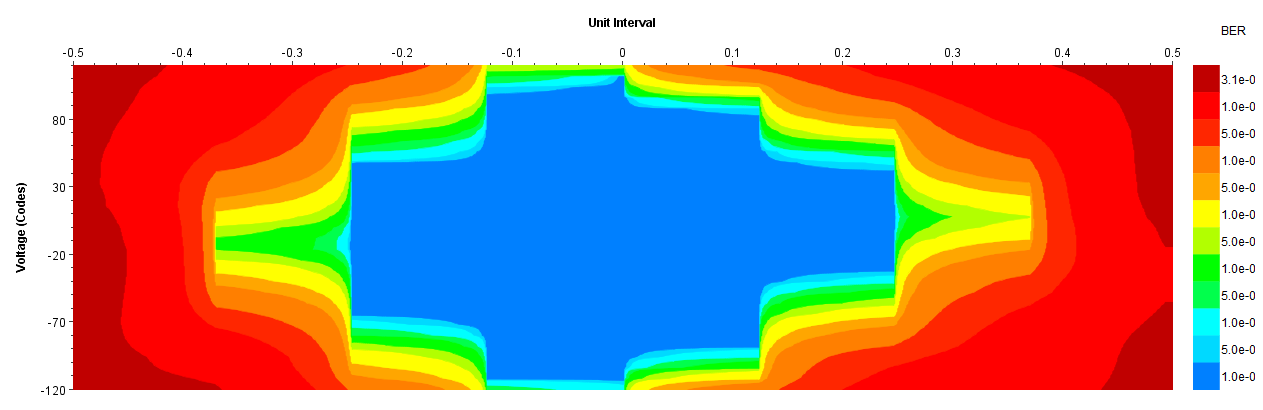}
        \caption{Slave lpGBT uplink:
        open area 5696 (\SI{56}{\percent}).}
        \label{fig:eyescan_slave}
    \end{subfigure}

    \caption{Representative two-dimensional BER eye scans of the
    \SI{10.24}{Gbps} CSM optical uplinks.
    The Vivado scans covered $-0.5$ to $+0.5$~UI in the horizontal
    direction and the full vertical sampling range, with horizontal and
    vertical increments of 8 and a dwell BER of \num{1e-5}. The color
    contours represent the BER measured as a function of receiver
    sampling phase and voltage offset.}
    \label{fig:lpgbt_eyescan}
\end{figure*}

In addition to the BER measurement at the nominal receiver sampling
point, two-dimensional BER eye scans were performed. The receiver sampling phase and
voltage offset were scanned while measuring the BER at each point.
Representative results for the master and slave uplinks are shown in
Fig.~\ref{fig:lpgbt_eyescan}. Both links show a wide open region,
demonstrating substantial timing and amplitude margins around the
nominal sampling point.

\begin{figure*}[t]
    \centering
    \includegraphics[width=\linewidth]{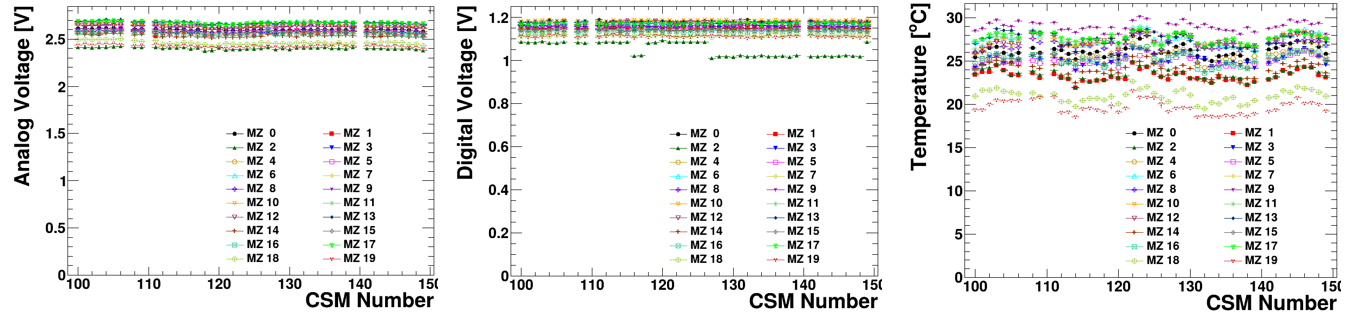}
    \caption{Measured analog voltage, digital voltage, and temperature from 20 mezzanine cards across multiple CSM units. Each point corresponds to one of the 31 GBT-SCA ADC channels read via the lpGBT slow control path. Consistent and stable readouts confirm the correct operation of the GBT-SCA monitoring system.}
    \label{fig:sca_readout}
\end{figure*}

The GBT-SCA--lpGBT monitoring path was evaluated by repeatedly reading
all ADC channels. Figure~\ref{fig:sca_readout} shows representative voltage and temperature readouts. Because
channel specific calibration constants were not applied, small
channel dependent gain and offset differences are expected. In particular, the point near \SI{1.0}{\volt} of digital voltage reflects channel dependent gain and offset in the uncalibrated ADC response. These
measurements are therefore used to verify stable and repeatable data
acquisition rather than absolute voltage accuracy. 

The JTAG chain was verified by reading the identifiers of all mezzanine cards. 
The FPGA fanout logic correctly routed the JTAG signals according to backend instructions.

All tested boards completed the validation sequence shown in Figure~\ref{fig:csm_test_flow} and met the defined acceptance criteria.

\subsection{Integration Tests}

A cosmic ray test was carried out at the University of Michigan to evaluate the CSM together with a small-diameter MDT (sMDT) chamber. For this purpose, a MiniDAQ system developed for MDT front-end electronics studies~\cite{guo2023development, guo2021design} was used. The system has been used in tests of the ASD, TDC, mezzanine cards, and CSM on sMDT chambers. It is compatible with the CSM optical uplink and slow control interfaces, and therefore provides a suitable setup for standalone data acquisition and system validation.

\begin{figure}[t]
\centering
\includegraphics[width=0.6\linewidth]{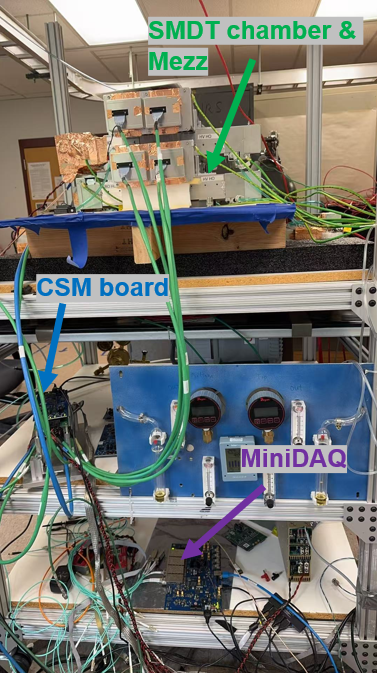}
\caption{Integration test platform with an sMDT chamber and the MiniDAQ system.}
\label{fig:integration_setup}
\end{figure}

As shown in Figure~\ref{fig:integration_setup}, the test bench includes a production CSM, mezzanine cards, an sMDT chamber, and scintillator paddles providing coincidence triggers for cosmic rays.

Performance benchmarks from the ATLAS Phase II sMDT upgrade show a spatial resolution of about \SI{80}{\um} at large drift radii, with an average value of \(99.9 \pm 7.0\)~\si{\um} measured across 30 modules~\cite{nelson2021performance}. These values are used here as a reference for the integration test and the subsequent analysis.

\begin{figure}[t]
\centering
\includegraphics[width=0.9\linewidth]{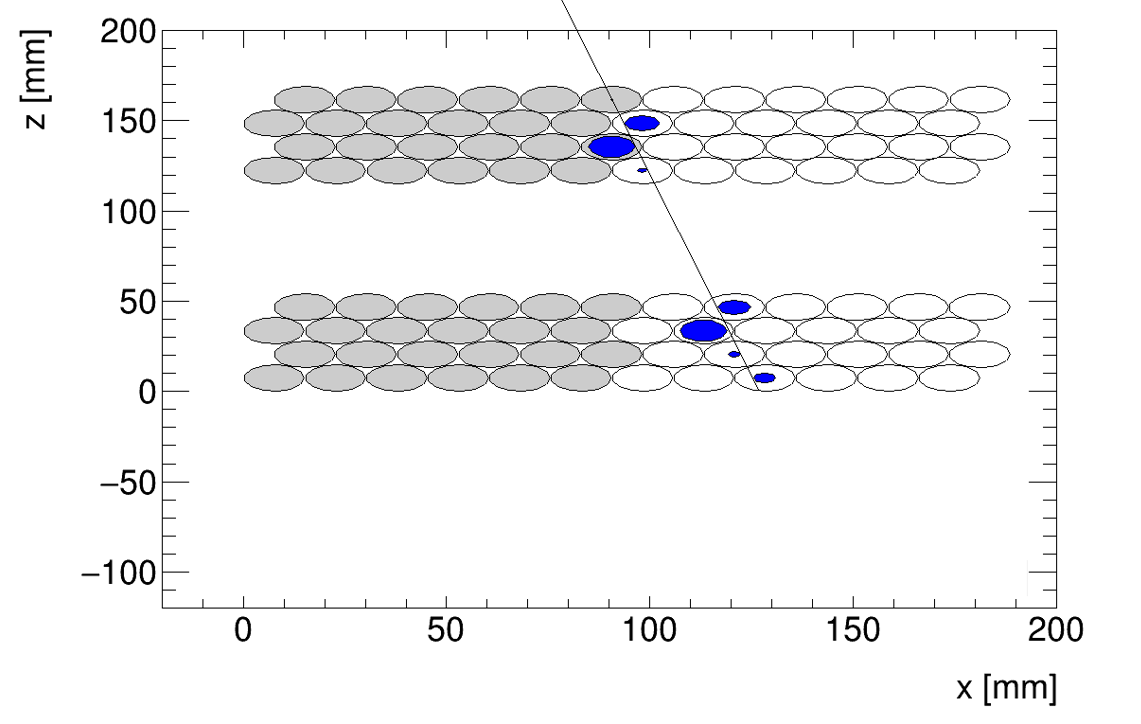}
\caption{Typical hitmap from cosmic ray data acquired using the sMDT chamber and CSM readout. Gray and white cell shading distinguishes groups of tubes read out by
different mezzanine cards. The blue ellipses indicate fired drift tubes for a representative muon track traversing the detector. The distribution demonstrates proper trigger logic, time digitization, and spatial correlation across layers.}
\label{fig:hitmap}
\end{figure}

Cosmic ray data were collected and used to reconstruct track segments from sMDT tube hits. One event is shown in Figure~\ref{fig:hitmap}. Residuals were calculated as the difference between measured hit positions and the fitted track. Two definitions were used: biased residuals, where the hit is included in the fit, and unbiased residuals, where it is excluded. The measured widths are \SI{79.5}{\um} and \SI{126.6}{\um}, respectively.

The spatial resolution was further studied as a function of drift radius. The drift radius was obtained from the TDC drift time using a pre-calibrated look up table that accounts for the non uniform electric field inside the tube. For each drift radius bin, biased and unbiased residual distributions were formed, and the resolution was calculated as the geometric mean:
$\sigma_\text{res} = \sqrt{\sigma_\text{biased} \cdot \sigma_\text{unbiased}}$~\cite{nelson2021performance}.

\begin{figure}[t]
\centering
\includegraphics[width=0.9\linewidth]{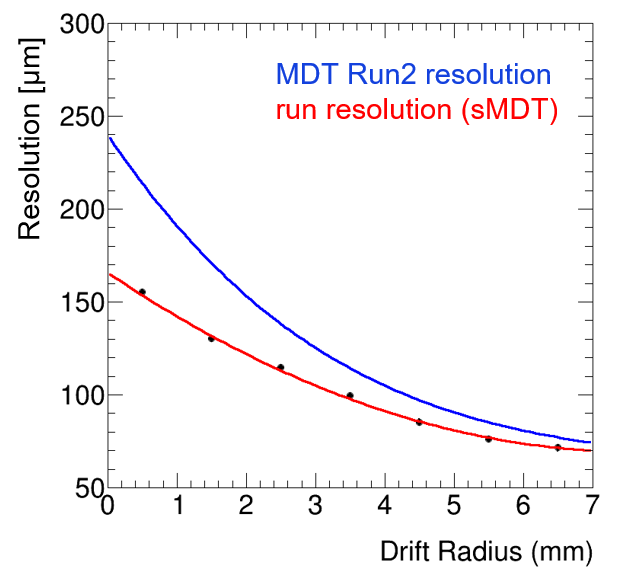}
\caption{Spatial resolution as a function of drift radius. The red
curve shows the result obtained with the
\SI{15}{\milli\meter}-diameter sMDT chamber used in the integration
test, while the blue curve shows the Run-2 MDT reference obtained with
\SI{30}{\milli\meter}-diameter tubes. The two curves therefore
correspond to different tube geometries and are presented as a
contextual performance comparison rather than as a direct
like-for-like comparison of the CSM electronics.}
\label{fig:resolution_curve}
\end{figure}

The resulting resolution profile is shown in Figure~\ref{fig:resolution_curve}. The measured resolution agrees with previous results from the construction phase. This indicates correct integration of the CSM with the sMDT readout chain.

To evaluate backend compatibility, four production CSM boards were tested at CERN and the Max Planck Institute (MPI) with a prototype L0MDT backend. The tests covered FPGA programming, configuration register access, and slow control operation. Table~\ref{tab:backend_tests} summarizes the test categories and the number of iterations.

\begin{table}[t]
\centering
\caption{Summary of backend integration tests with L0MDT prototype.}
\label{tab:backend_tests}
\begin{tabular}{@{}p{5cm}p{2cm}@{}}
\toprule
\textbf{Test Type} & \textbf{Iterations}\\
\midrule
lpGBT register write-readback  & 1,000/LpGBT \\
GBT-SCA ID readout per board  & 1,000/CSM \\
Environmental monitoring readings & 10,000/CSM  \\
\bottomrule
\end{tabular}
\end{table}

As part of the test procedure for each CSM, 1,000 register
write-readback iterations were performed on each of the two lpGBT
ASICs. No mismatches or communication errors were observed.

The identification registers of the four GBT-SCA chips were also read out. Each CSM was queried 1,000 times, and all responses were consistent.

Environmental monitoring was evaluated through 10,000 repeated
readouts from the GBT-SCA ADC channels. Because channel-specific
calibration constants were not applied in this test, the resulting
voltage and temperature values are presented as representative
uncalibrated distributions rather than as absolute-accuracy
measurements. The test was used to verify the stability and
repeatability of the monitoring path, and no discontinuities,
abnormal temporal variations, communication errors, or data loss were
observed.

\subsection{Thermal Stress Tests}

Thermal qualification included IPC D-coupon tests and board-level
thermal cycling of assembled CSMs. The IPC thermal shock and reflow
tests were performed on representative D-coupons fabricated with the
same PCB stack up and manufacturing process as the CSM. Assembled
boards intended for detector deployment were not subjected to these
accelerated IPC tests.

Twelve D-coupons were subjected to the thermal-shock test following
IPC-TM-650. A total of 100 cycles were performed between
\SI{-55}{\celsius} and \SI{180}{\celsius}, with a dwell time of
\SI{15}{\minute} at each temperature and a transition time below
\SI{2}{\minute}. Coupon resistance was measured using a four-wire
Kelvin bridge and compared with the pre-test baseline. The acceptance criterion was a resistance
variation below \SI{5}{\percent}. All 12 coupons satisfied this
requirement.

A separate set of 12 D-coupons was subjected to six thermal reflow
cycles following IPC-TM-650, with each cycle reaching a peak
temperature of \SI{260}{\celsius}. Resistance was measured relative to
the pre-reflow baseline using the same four-wire method and
\SI{5}{\percent} acceptance criterion. All 12 coupons satisfied the
requirement.

Separately, all 50 pre-production CSM boards underwent four moderate
board-level thermal cycles. Each cycle consisted of
\SI{100}{\minute} at \SI{55}{\celsius}, followed by a
\SI{20}{\minute} recovery at room temperature, with a temperature
ramp rate of \SI{10}{\celsius\per\minute}. In addition to verifying
functional stability under repeated temperature excursions, this
procedure served as a mild burn-in screen for latent assembly or
component defects that could produce early-life failures. After
thermal cycling, the production-level functional tests were repeated
for every board, and no degradation relative to the pre-cycling
results was observed. This thermal-screening procedure will be
included in the quality-assurance program for all production batches.

\subsection{Radiation Tolerance Tests}

Radiation tolerance was checked for the key components (Artix-7 FPGA and Texas Instruments TPS7A8500 LDO \cite{ti_tps7a85}) used in the production CSM. The tests followed the radiation studies reported in Ref.~\cite{hu2026radiation}, where the FPGA was evaluated for Total Ionizing Dose (TID) and Single Event Effects (SEE), and the LDO regulators were evaluated for TID, SEE, and Non-Ionizing Energy Loss (NIEL). During the TID and SEE irradiation tests, the DUT boards were powered and operated. The FPGA exercised the fanout path using an \SI{80}{Mbps} PRBS-31 stream, while the LDO output voltages and currents were monitored. The referenced irradiation study does not report an additional representative LDO load during the TID or SEE exposure. During the NIEL irradiation, the LDO boards were unpowered with their power leads grounded. Seven days after irradiation, the LDO outputs were evaluated using a \SI{3.3}{\volt} input and a \SI{2.2}{\ohm} load. The CSM does not employ an on-board nonvolatile memory for FPGA configuration. Therefore, no separate nonvolatile FPGA configuration device requires radiation qualification. 

Table~\ref{tab:radiation_rtc_summary} summarizes the nominal
10-year radiation exposure corresponding to
\SI{4000}{\femto\barn^{-1}}, the fully safety-factored ATLAS COTS
Radiation Tolerance Criteria (RTC), and the corresponding irradiation
conditions. The nominal exposure values are given before application
of the RTC safety factors. The COTS RTC values additionally include
the simulation uncertainty factor and the radiation-effect-specific
and COTS safety factors~\cite{hu2026radiation}. The TID tests provide a direct margin relative to the RTC dose. The SEE qualification is based on the measured functional error cross-section and the projected error rate under the ATLAS RTC hadron fluence. For the LDO NIEL test, the received fluence depended on the device position, and the post-irradiation output voltages were used to evaluate the device performance. 

\begin{table*}[t]
    \centering
    \caption{Summary of the nominal 10-year radiation exposure, the
    fully safety-factored ATLAS COTS Radiation Tolerance Criteria (RTC),
    and the irradiation conditions used for the COTS components on the
    CSM. The nominal exposure corresponds to
    \SI{4000}{\femto\barn^{-1}} before application of the RTC safety
    factors.}
    \label{tab:radiation_rtc_summary}

    \small
    \setlength{\tabcolsep}{5pt}
    \renewcommand{\arraystretch}{1.25}

    \begin{tabularx}{\textwidth}{
        @{}
        l
        l
        >{\raggedright\arraybackslash}p{0.27\textwidth}
        >{\raggedright\arraybackslash}X
        @{}
    }
        \hline
        \textbf{Effect} &
        \textbf{Component} &
        \textbf{Radiation level} &
        \textbf{Test condition / exposure} \\
        \hline

        TID &
        FPGA &
        \begin{tabular}[t]{@{}l@{}}
        Nominal 10-year: \(2.1~\mathrm{krad}\) \\
        COTS RTC: \(9.6~\mathrm{krad}\)
        \end{tabular} &
        \(^{60}\mathrm{Co}\) irradiation at BNL, the FPGA remained
        functional up to approximately \(200~\mathrm{krad}\), with one
        FPGA-only test reaching failure at approximately
        \(626~\mathrm{krad}\). \\

        TID &
        LDO &
        \begin{tabular}[t]{@{}l@{}}
        Nominal 10-year: \(2.1~\mathrm{krad}\) \\
        COTS RTC: \(9.6~\mathrm{krad}\)
        \end{tabular} &
        \(^{60}\mathrm{Co}\) irradiation at BNL, the LDOs remained
        operational after approximately \(92~\mathrm{krad}\), followed
        by annealing. \\

        SEE &
        FPGA &
        \begin{tabular}[t]{@{}l@{}}
        Nominal 10-year: \(2.1\times10^{11}~h/\mathrm{cm}^{2}\) \\
        COTS RTC: \(19.0\times10^{11}~h/\mathrm{cm}^{2}\) \\
        \(E>20~\mathrm{MeV}\)
        \end{tabular} &
        Neutron irradiation at LANSCE, functional-error cross sections
        were extracted from the observed soft and hard errors and used
        to estimate the error rate under the RTC fluence. \\

        SEE &
        LDO &
        \begin{tabular}[t]{@{}l@{}}
        Nominal 10-year: \(2.1\times10^{11}~h/\mathrm{cm}^{2}\) \\
        COTS RTC: \(19.0\times10^{11}~h/\mathrm{cm}^{2}\) \\
        \(E>20~\mathrm{MeV}\)
        \end{tabular} &
        Neutron irradiation at LANSCE together with the FPGA SEE test,
        the LDO output voltages were monitored during irradiation. \\

        NIEL &
        LDO &
        \begin{tabular}[t]{@{}l@{}}
        Nominal 10-year:
        \(11.6\times10^{11}~n_{\mathrm{eq}}/\mathrm{cm}^{2}\) \\
        COTS RTC:
        \(68.1\times10^{11}~n_{\mathrm{eq}}/\mathrm{cm}^{2}\)
        \end{tabular} &
        Irradiation at the Frascati Neutron Generator, the received
        1-MeV-equivalent neutron fluence ranged from
        \(2.3\times10^{12}\) to
        \(5.6\times10^{12}~n_{\mathrm{eq}}/\mathrm{cm}^{2}\),
        depending on device position. \\

        NIEL &
        FPGA &
        \begin{tabular}[t]{@{}l@{}}
        Nominal 10-year:
        \(11.6\times10^{11}~n_{\mathrm{eq}}/\mathrm{cm}^{2}\) \\
        COTS RTC:
        \(68.1\times10^{11}~n_{\mathrm{eq}}/\mathrm{cm}^{2}\)
        \end{tabular} &
        Not tested, NIEL testing was omitted because displacement
        damage is not expected to be a limiting effect for modern CMOS
        devices~\cite{hu2026radiation}. \\

        \hline
    \end{tabularx}
\end{table*}




The TID test was performed at the \(^{60}\mathrm{Co}\) facility at Brookhaven National Laboratory (BNL). For the FPGA tests, no FPGA failure was observed up to about \SI{200}{krad}. In these runs, the loss of board power was caused by the on-board LDOs rather than by the FPGA itself. A separate FPGA-only test was therefore performed with the LDOs shielded, and the FPGA failed only after a dose of about \SI{626}{krad}. The LDO regulators were tested to about \SI{92}{krad} without functional failure. End-of-life failures were observed only at higher doses, between \SI{150}{krad} and \SI{200}{krad}. The irradiated samples were annealed and re-tested after the TID exposure, with no abnormal recovery behavior reported in Ref.~\cite{hu2026radiation}.

The SEE test was carried out at the ICE-II facility of the Los Alamos
Neutron Science Center (LANSCE), using a broad-spectrum neutron beam
extending to approximately \SI{800}{\mega\electronvolt}. FPGA errors
were classified as soft if normal operation could be restored without
power cycling, and as hard if a power cycle was required. Soft-error
recovery includes reset, SEM-based configuration-memory correction, and
remote FPGA reconfiguration, with a typical recovery time of
approximately \SI{10}{\second}. Hard errors require board power cycling
and approximately \SI{5}{\minute} for recovery, including operator
intervention~\cite{hu2026radiation}.

Using the worst-case measured functional-error cross section of
approximately \qty{2.4e-11}{\centi\meter\squared}, the expected error
rate was evaluated with the ATLAS COTS RTC fluence of
\qty{1.90e12}{h/\cm^2}. This fluence corresponds to the projected
10-year HL-LHC exposure and already includes the prescribed RTC safety
factors. The resulting conservative rate is approximately one
functional error every \SI{2.5}{month} per FPGA. When the same conservative per-device rate is scaled to the full
population of 1,104 CSM FPGAs, it corresponds to approximately
442 functional errors per month, or about 14.5 events per day across
the detector. This value should not be interpreted as a simultaneous loss of the full system. Each event is localized to the affected FPGA and can be handled independently through the adopted SEM based mitigation and recovery procedure.

The NIEL response of the LDO regulators was evaluated at the Frascati
Neutron Generator. Depending on their positions in the irradiation
setup, the devices received \SI{1}{MeV}-equivalent neutron fluences
between \(\num{2.3e12}\) and
\(\qty{5.6e12}{n_{\mathrm{eq}}/\cm^2}\). Measurements before and after irradiation showed no measurable shift in the output voltages relative to their corresponding pre-irradiation
values. For comparison, the
maximum tested fluence is about 4.8 times the nominal 10-year exposure
of \(\qty{11.6e11}{n_{\mathrm{eq}}/\cm^2}\), and reaches 82\% of the
fully safety-factored COTS RTC value of
\(\qty{68.1e11}{n_{\mathrm{eq}}/\cm^2}\). The FPGA was not included in
the NIEL campaign because displacement damage is not expected to be a
limiting effect for modern CMOS devices~\cite{hu2026radiation}.

The TID and SEE measurements demonstrate that the tested FPGA and LDO
devices satisfy the corresponding ATLAS radiation-tolerance criteria.
For NIEL, the LDO showed no measurable degradation up to the maximum
tested fluence of
\(\qty{5.6e12}{n_{\mathrm{eq}}/\cm^2}\). The qualification is therefore
stated up to this tested fluence.

Radiation-hard ASICs used in the design, including the lpGBT, VTRx+, and GBT-SCA, have been qualified separately and are not discussed in this section.

\section{Conclusion}

The Phase-II CSM provides the high-bandwidth readout, timing distribution, slow control, and remote configuration functions required for the upgraded ATLAS MDT system. The design has been validated through production-level electrical and link tests, integration with sMDT chambers and the prototype L0MDT backend, thermal qualification, and radiation tests of the selected COTS components. The TID and SEE tests satisfy the corresponding radiation-tolerance
criteria, while the LDO NIEL qualification is demonstrated up to the
maximum tested fluence of
\(\qty{5.6e12}{n_{\mathrm{eq}}/\cm^2}\), at which no measurable
degradation was observed. The measured results demonstrate the required electrical, data-transmission, integration, and thermal performance of the CSM. The production CSM boards are being prepared for installation and integration during the upcoming LHC long shutdown. 

\section{Acknowledgment}

The work is supported by the US National Science Foundation (NSF) and the US Department of Energy (DOE) under contracts PHY1948993 (NSF) and DE-SC007859 (DOE). 

\bibliographystyle{unsrt}

\bibliography{cas-refs}


\appendix
\section{Appendix: Data Format}\label{appendix}

In the following tables, FRMUP and FRSUP denote the bit positions of
the master and slave lpGBT uplink frames, respectively, while FRMDN
denotes the bit positions of the master lpGBT downlink frame. FEC
denotes forward error correction, EC and IC denote the external- and
internal-control fields, respectively, H denotes the frame header, and
DownIC denotes the downlink IC field.

\setcounter{table}{0}
\renewcommand{\thetable}{\Alph{section}\arabic{table}}

\begin{table}[H]
\centering
\caption{Master lpGBT uplink frame (FRMUP) mapping in the CSM
prototype (FEC5 mode).}
\label{tab:master_uplink}
\scriptsize
\begin{tabular}{llll}
\hline
Uplink frame (FRMUP) & Field & CSM data mapping \\
\hline
FRMUP[19:0]   & FEC[19:0]         & N/A         \\

FRMUP[27:20]  & Data[7:0]          & TDC6 CH1    \\
FRMUP[35:28]  & Data[15:8]         & TDC6 CH2    \\
FRMUP[43:36]  & Data[23:16]        & TDC8 CH1    \\
FRMUP[51:44]  & Data[31:24]        & TDC8 CH2    \\

FRMUP[59:52]  & Data[39:32]        & TDC5 CH1    \\
FRMUP[67:60]  & Data[47:40]        & TDC5 CH2    \\
FRMUP[75:68]  & Data[55:48]        & TDC7 CH1    \\
FRMUP[83:76]  & Data[63:56]        & TDC7 CH2    \\

FRMUP[91:84]  & Data[71:64]        & TDC9 CH1    \\
FRMUP[99:92]  & Data[79:72]        & TDC9 CH2   \\
FRMUP[107:100]& Data[87:80]                & SCA1 E-link AUX \\
FRMUP[115:108]& Data[95:88]                & SCA1 E-link PRI \\

FRMUP[123:116]& Data[103:96]               & SCA2 E-link AUX \\
FRMUP[131:124]& Data[111:104]              & SCA2 E-link PRI \\
FRMUP[139:132]& Data[119:112]              & SCA3 E-link AUX \\
FRMUP[147:140]& Data[127:120]              & SCA3 E-link PRI \\

FRMUP[155:148]& Data[135:128]              & SCA0 E-link AUX \\
FRMUP[163:156]& Data[143:136]              & SCA0 E-link PRI \\
FRMUP[171:164]& Data[151:144]      & TDC3 CH2    \\
FRMUP[179:172]& Data[159:152]      & TDC3 CH1    \\

FRMUP[187:180]& Data[167:160]      & TDC1 CH2    \\
FRMUP[195:188]& Data[175:168]      & TDC1 CH1    \\
FRMUP[203:196]& Data[183:176]      & TDC4 CH2    \\
FRMUP[211:204]& Data[191:184]      & TDC4 CH1    \\

FRMUP[219:212]& Data[199:192]      & TDC2 CH2    \\
FRMUP[227:220]& Data[207:200]      & TDC2 CH1    \\
FRMUP[235:228]& Data[215:208]      & TDC0 CH2    \\
FRMUP[243:236]& Data[223:216]      & TDC0 CH1    \\

FRMUP[249:244]& \{4'b0, DownIC[1:0]\} & N/A   \\
FRMUP[251:250]& EC[1:0]             & S-lpGBT cfg \\
FRMUP[253:252]& IC[1:0]             & M-lpGBT cfg \\
FRMUP[255:254]& H[1:0]            & Header   \\
\hline
\end{tabular}
\end{table}

\begin{table}[p]
\centering
\caption{Slave lpGBT uplink frame (FRSUP) mapping in the CSM
prototype (FEC5 mode).}
\label{tab:slave_uplink}
\scriptsize
\begin{tabular}{llll}
\hline
Uplink frame (FRSUP) & Field & CSM data mapping \\
\hline
FRSUP[19:0]   & FEC[19:0]         & N/A         \\

FRSUP[27:20]  & Data[7:0]          & TDC16 CH1   \\
FRSUP[35:28]  & Data[15:8]         & TDC16 CH2   \\
FRSUP[43:36]  & Data[23:16]        & TDC18 CH1   \\
FRSUP[51:44]  & Data[31:24]        & TDC18 CH2   \\

FRSUP[59:52]  & Data[39:32]        & TDC15 CH1   \\
FRSUP[67:60]  & Data[47:40]        & TDC15 CH2   \\
FRSUP[75:68]  & Data[55:48]        & TDC17 CH1   \\
FRSUP[83:76]  & Data[63:56]        & TDC17 CH2   \\

FRSUP[91:84]  & Data[71:64]        & TDC19 CH1   \\
FRSUP[99:92]  & Data[79:72]        & TDC19 CH2   \\
FRSUP[107:100]& Data[87:80]        & N/A         \\
FRSUP[115:108]& Data[95:88]        & N/A         \\

FRSUP[123:116]& Data[103:96]       & N/A         \\
FRSUP[131:124]& Data[111:104]      & N/A         \\
FRSUP[139:132]& Data[119:112]      & TDC13 CH2   \\
FRSUP[147:140]& Data[127:120]      & TDC13 CH1   \\

FRSUP[155:148]& Data[135:128]      & TDC11 CH2   \\
FRSUP[163:156]& Data[143:136]      & TDC11 CH1   \\
FRSUP[171:164]& Data[151:144]      & TDC14 CH2   \\
FRSUP[179:172]& Data[159:152]      & TDC14 CH1   \\

FRSUP[187:180]& Data[167:160]      & N/A         \\
FRSUP[195:188]& Data[175:168]      & N/A         \\
FRSUP[203:196]& Data[183:176]      & N/A         \\
FRSUP[211:204]& Data[191:184]      & N/A         \\

FRSUP[219:212]& Data[199:192]      & TDC12 CH2   \\
FRSUP[227:220]& Data[207:200]      & TDC12 CH1   \\
FRSUP[235:228]& Data[215:208]      & TDC10 CH2   \\
FRSUP[243:236]& Data[223:216]      & TDC10 CH1   \\

FRSUP[249:244]& \{4'b0, DownIC[1:0]\}  & N/A  \\
FRSUP[251:250]& EC[1:0]             & \\
FRSUP[253:252]& IC[1:0]             &  \\
FRSUP[255:254]& H[1:0]            & Header   \\
\hline
\end{tabular}
\end{table}

\begin{table}[p]
\centering
\caption{Master lpGBT downlink frame mapping in the CSM prototype (2.56\,Gbps).}
\label{tab:downlink_format}
\scriptsize
\begin{tabular}{llll}
\hline
Frame & Function  & Control E-link Mapping \\
\hline
FRMDN[23:0]   & FEC[23:0]         & N/A \\

FRMDN[25:24]  & Data[1:0]            & ENC \\
FRMDN[27:26]  & Data[3:2]            & N/A \\
FRMDN[29:28]  & Data[5:4]            & ENC (Backup) \\
FRMDN[31:30]  & Data[7:6]            & N/A \\

FRMDN[33:32]  & Data[9:8]            & N/A \\
FRMDN[35:34]  & Data[11:10]          & SCA1 E-link PRI \\
FRMDN[37:36]  & Data[13:12]          & SCA1 E-link AUX \\
FRMDN[39:38]  & Data[15:14]          & SCA2 E-link AUX \\

FRMDN[41:40]  & Data[17:16]          & SCA2 E-link PRI \\
FRMDN[43:42]  & Data[19:18]          & SCA3 E-link AUX \\
FRMDN[45:44]  & Data[21:20]          & N/A \\
FRMDN[47:46]  & Data[23:22]          & SCA3 E-link PRI \\

FRMDN[49:48]  & Data[25:24]          & SCA0 E-link AUX \\
FRMDN[51:50]  & Data[27:26]          & SCA0 E-link PRI \\
FRMDN[53:52]  & Data[29:28]          & N/A \\
FRMDN[55:54]  & Data[31:30]          & N/A \\

FRMDN[57:56]  & EC[1:0]              & S-lpGBT cfg \\
FRMDN[59:58]  & IC[1:0]              & M-lpGBT cfg \\

FRMDN[63:60]  & H[3:0]           & Header  \\
\hline
\end{tabular}
\end{table}




\end{document}